# On non-canonical solving the Satisfiability problem

**Anatoly D. Plotnikov**

Department of Social Informatics and Safety of Information Systems,
Dalh East-Ukrainian National University,
Luhansk, Ukraine
Email: a.plotnikov@list.ru

### Abstract

We study the non-canonical method for solving the Satisfiability problem which given by a formula in the form of the conjunctive normal form. The essence of this method consists in counting the number of tuples of Boolean variables, on which at least one clause of the given formula is false. On this basis the solution of the problem obtains in the form YES or NO without constructing tuple, when the answer is YES. It is found that if the clause in the given formula has pairwise contrary literals, then the problem can be solved efficiently. However, when in the formula there are a long chain of clauses with pairwise non-contrary literals, the solution leads to an exponential calculations.

## 1 Introduction

To uniquely understand used terminology in the article, we recall some definitions.

Suppose we have a two-element set $E = \{0;1\}$. The function $f: E^n \to E$ is called *Boolean*. Every vector in $E^n$ is considered as the tuple of values of variables $x_1, x_2, ..., x_n$ of the function $f(x_1, x_2, ..., x_n)$.

A Boolean variable with negation or without negation is called *literal*. Literals $x$ and $\bar{x}$ is called *contrary*. The conjunction of $r$ $(1 \leq r \leq n)$ different non-contrary literals, called *elementary*. Any Boolean function can be represented as a disjunction of some elementary conjunctions. Such a representation of the function is called *disjunctive normal form (DNF)* of the function $f(x_1, x_2, ..., x_n)$. For example, the function can be written as the following DNF:

$$f(x_1, x_2, x_3) = x_1 \vee x_2 x_3 \vee \overline{x_2} \overline{x_3}.$$

A disjunction of different non-contrary literals called *clause*. Any Boolean function can be represented as a conjunction of some clauses. Such a representation of the function is called *conjunctive normal form (CNF)* of the $f(x_1, x_2, ..., x_n)$ can be written as CNF:

$$f(x_1, x_2, x_3) = (x_1 \vee x_2 \vee \overline{x_3})(x_1 \vee \overline{x_2} \vee x_3).$$

In the theory of computational complexity, the problems, belonging to the class NP, are considered. The problem Z belongs to the class NP if its solution can be verified by the time, described of a polinomial function on the dimension of the problem. Usually, a problem of the class NP formulated as a decision problem, that is, in the form of a question to which the answer (solution) is YES or NO. Thus, the solution of the problem — it is the answer YES or NO [2]. To verify the correctness of the answer to this question, some certificate should be submitted that in polynomial time on the dimension of the problem verifies the correctness of the answer.

However, in practice, the solution of the problem in form of a set of some parameters of the problem (vertices, edges, literals) is used as a certificate, allowing form the answer YES or NO. To obtain the certificate in such form is often difficult [11].

Of particular importance in the theory of computational complexity plays the satisfiability problem, which is formulated as follows [1, 2]. Let it is given $m$ clauses $D_1, D_2, ..., D_m$, containing variables $x_1, x_2, ..., x_n$. Has the formula $f(x_1, x_2, ..., x_n) = D_1 \cdot D_2 \cdot ... \cdot D_m$ satisfiable assignment?

The SAT problem is one of NP-complete problems [1-10]. Historically, this is the first NP-complete problem. Solution of this problem is important for theory and practice. Currently, to solve the SAT problem, DPLL algorithms found the greatest application which are based on the works of Davis and Putnam [12] and Davis Lodzhmana and Loveland [13]. The essence of these algorithms is to find a satisfying assignment for the SAT problem. The complexity of these algorithms can be described by an exponential formula of the dimension of the problem, since it is related to the construction of the tree of values of the variables of the problem [14].

The solution of the Satisfiability problem is called non-canonical if the answer YES or NO was obtained without presentation of variable values when a given formula is true.

The purpose of this article is to investigate of non-canonical method for solving the Satisfiability problem.

## 2 Basis of the method

Let the formula of the Satisfiability problem is:

$$f(x_1, x_2, ..., x_n) = D_1 \cdot D_2 \cdot ... \cdot D_m. \qquad (1)$$

It is necessary to determine the existence of unknown values $x_1, x_2, ..., x_n$ such that satisfy the function (1).

The formula (1) can be transformed by the operation of negation on it. As a result, we will obtain a new formula:

$$\overline{f}(x_1, x_2, ..., x_n) = C_1 \vee C_2 \vee ... \vee C_m, \qquad (2)$$

where each elementary conjunction obtained by the negation of the corresponding clause: $C_i = \overline{D_i}$ ($i = 1, 2, ..., m$) in the formula (1).

**Lemma 1.** *Each conjunction in the formula (2) defines a set of tuples of variables values of the formula (1), for which the corresponding clause is false.*

**Proof.** In fact, in the formula (1), any clause is false if and only if every literal of this clause is false. Therefore, an elementary conjunction of the formula (2) must be true for the tuples on which the corresponding clause is false. What we wanted to prove. ◊

For example, suppose we have the formula:

$$f(x_1, x_2, x_3) = (x_1 \vee \overline{x_2})(x_2 \vee \overline{x_3})(\overline{x_1} \vee x_3), \qquad (3)$$

where $D_1 = x_1 \vee \overline{x_2}$, $D_2 = x_2 \vee \overline{x_3}$, $D_3 = \overline{x_1} \vee x_3$.

The truth table for the clauses of this formula has the form:

Table 1: Table for the clauses

| $x_1$ | $x_2$ | $x_3$ | $D_1$ | $D_2$ | $D_3$ |
|---|---|---|---|---|---|
| 0 | 0 | 0 | 1 | 1 | 1 |
| 0 | 0 | 1 | 1 | 0 | 1 |
| 0 | 1 | 0 | 0 | 1 | 1 |
| 0 | 1 | 1 | 0 | 1 | 1 |
| 1 | 0 | 0 | 1 | 1 | 0 |
| 1 | 0 | 1 | 1 | 0 | 1 |
| 1 | 1 | 0 | 1 | 1 | 0 |
| 1 | 1 | 1 | 1 | 1 | 1 |

Executing a negation operation on the formula (3), we obtain the formula:
$$\overline{f(x_1, x_2, x_3)} = \overline{x_1} \cdot x_2 \vee \overline{x_2} \cdot x_3 \vee x_1 \cdot \overline{x_3}. \quad (4)$$

Thus, we conclude that the set of elementary conjunctions of the formula (2) represents the set of all tuples of variables $x_1, x_2, ..., x_n$, on which at least one clause of formula (1) is false.

**Theorem 1.** *The Satisfiable problem, defined on a set of variables has a satisfiable assignment if and only if the total number of different tuples, on which at least one clause in the formula (1) is false, strictly smaller than $2^n$.*

**Proof.** This is evident. If the total number of tuples, in which at least one clause of formulas (1) is false, is equal to $2^n$, then the satisfiable assignment does not exist for the Satisfiability problem. ◊

## 3 Certificate

Thus, according to Theorem 1 if you calculate the total number of different tuples, in which at least one clause in the Satisfiability problem is false, then on this basis, we can formulate the answer YES or NO to the question about existence of the satisfying assignment, that is, one can solve the Satisfiability problem, without constructing satisfying assignment. In this case, the certificate counts the total number of different tuples, in which at least one clause of the formula (1) is false.

Consider the different cases of the counting of tuples, in which clause of the formula (1) is false.

**Lemma 2.** *Each conjunction $C_i$ $(i = 1, 2, ..., m)$ of the equation (2), containing $r_i$ ($1 \leq r_i \leq n$) literals, defines $2^{n-r_i}$ tuples on which the corresponding clause of the formula (1) is false.*

**Proof.** Let us consider the elementary conjunction in the formula (2). Let $C_i$ $(i = 1, 2, ..., m)$ is one of them. If the number of literals that make up the $C_i$ is $r_i$ ($1 \leq r_i \leq n$), then this conjunction defines $2^{n-r_i}$ of tuples on which the corresponding clause of the formula (1) is false, because the $(n - r_i)$ missing literals can take all possible values. ◊

So, in the example (see Table 1 above), conjunction $C_1 = \overline{x_1} \cdot x_2$ determines $2^{3-2} = 2$ of tuples on which clause $D_1 = x_1 \vee \overline{x_2}$ is false. Here we have $n = 3$; $r_i = 2$. These are tuples (010) and (011).

**Lemma 3.** *Any pair of elementary conjunctions $C_i$ and $C_k$ $(i, k = 1, 2, ..., m; i \neq k)$ of formula (2), containing contrary literals together, determines $2^{n-r_i} + 2^{n-r_k}$ of tuples on which the corresponding clause of the formula (1) is false.*

**Proof.** Let us now consider two elementary conjunction of the formula (2). Let it be the conjunctions of the $C_i$ and $C_k$ $(i, k = 1, 2, ..., m; i \neq k)$. Obviously, if the literal of the $C_i$ has the contrary literal in $C_k$, then these conjunctions determine different tuples of variables. ◊

For example, two elementary conjunctions in the formula (4) $C_1 = \overline{x_1} \cdot x_2$ and $C_3 = x_1 \cdot \overline{x_3}$ ($n = 3$) contain the variable $x_1$ with negation and without negation. The conjunction $C_3$ determine tuples (100) and (110), which does not coincide with the tuples, represented by the conjunction $C_1$ ((010), (011)).

The considered conjunctions are called *alternative*.

**Theorem 2.** *Let the formula (2) contains only pairwise alternative conjunctions $C_i, C_k$ $(i, k = 1, 2, ..., m; i \neq k)$. Then the total number of tuples, on which the corresponding clause of the formula (1) is false, is equal*

$$\sum_{i=1}^{m} 2^{n-r_i}.$$

**Proof.** It follows from Lemma 3. ◊

It is easy to see that the formula (4) contains pairwise alternative conjunctions. The total number of tuples, on which the corresponding clauses in the formulas (3) is false, is equal:
$$2^1 + 2^1 + 2^1 = 6,$$
that is, the formula (3) has $2^3 - 6 = 2$ satisfiable tuples. In this case, the solution of the problem is YES.

**Theorem 3.** *Any Boolean function, represented by the formula (2), can be transformed into a formula containing only pairwise alternative conjunctions.*

**Proof.** Any Boolean function can be represented as a perfect disjunctive normal form (PDNF), which contains only pairwise alternative conjunctions, each of which depends on *n* variables. For example, a pair of conjunctions $x_1 x_2 \lor x_1 x_3$ can be transformed as follows:

$$x_1 x_2 \lor x_1 x_3 = x_1 x_2 (x_3 \lor \overline{x_3}) \lor x_1 (x_2 \lor \overline{x_2}) x_3 = x_1 x_2 x_3 \lor x_1 x_2 \overline{x_3} \lor x_1 \overline{x_2} x_3. \diamond$$

Unfortunately the time (the number of operations), that is required to build PDNF, can be unacceptably large.

Let there is some conjunction $C_i$ containing $r_i$ literals from *n* ( $r_i \leq n$). The variables, which do not have their own literals in $C_i$, called *free* for the given conjunction. The number of such free variables in the conjunction $C_i$ equals $t_i = n - r_i$. Similarly, if there are *p* conjunctions $C_1, C_2, ..., C_p$, then the total number of free variables in these conjunctions is designated as $t_{1,2,...,p}$.

**Theorem 4.** *Let the formula (2) has p conjunctions $C_1, C_2, ..., C_p$, having no pairwise alternative literals. Then the total number of tuples, on which this conjunction is false, is equal to:*

$$R = \sum_{i=1}^{p} 2^{t_i} - \sum_{\forall i < j} 2^{t_{i,j}} + \sum_{\forall i < j < k} 2^{t_{i,j,k}} - ... + (-1)^{p-1} 2^{t_{1,2,...,p}} \quad (5)$$

**Proof.** The validity of Theorem 4 follows from the method of inclusions and exclusions. ◊

Consider a chain of pairwise non-alternative conjunctions $C_1 = x_1$, $C_2 = x_2$, $C_3 = x_3 \cdot x_4$. The total number of tuples, on which the corresponding disjunctions are false, is equal to:
$$R = 2^3 + 2^3 + 2^2 - 2^2 - 2^1 - 2^1 + 2^0 = 13.$$

**Theorem 5.** *Let the formula (2) has p conjunctions $C_1, C_2, ..., C_p$, having no pairwise alternative literals. Then the number of members in the formula (5) is equal to $2^{p-1}$.*

**Proof.** The number of members in the formula (5) equals the number of subsets of a fnite p-set without empty subset. ◊

## 4 Conclusions

Consideration of non-canonical method for solving the Satisfiability problem shows that this method can be used effectively for solving problems, initially containing pairwise of alternative conjunctions. Unfortunately, in the general case, solving the problem by the considered method is ineffectively as it involves performing an exponential number of operations. Therefore, this method can be used with a limited number of chains of pairwise non-alternative conjunctions. Obviously, this depends on the performance of the used computer.

## References

[1] Garey M.R. and Johnson D.S. (1979) *Computers and Intractability.* W.H. Freeman and Company, San Francisco.

[2] Papadimitriou C.H. and Steiglitz K. (1982) *Combinatorial optimization: Algorithms and complexity.* Prentice-Hall Inc., New Jersey.

[3] Cook S. (1971) The Complexity of Theorem-Proving Procedures. *Proceedings of the 3rd Annual ACM Symposium on Theory of Computing*: pp. 151–158.

[4] Karp R.M. (1972) Reducibility among combinatorial problems. In Miller, Raymond E.: Thatcher, James W. *Complexity of Computer Computations*. Plenum. pp. 85–103.

[5] Crescenzi P. and Kann V. (1997) Approximation on the Web: A Compendium of NP Optimization Problems. *Proceedings of International Workshop RANDOM'97*, Bologna, 11-12 July 1997, pp. 111-118.

[6] Dunne P.E. (2008) An annotated list of selected NP-complete problems. COMP202, Dept. of Computer Science, University of Liverpool.

[7] Levin L.A. (1973) Universal Searching Problems. *Prob. Info. Transm.* **9**, 265-266.

[8] Valiant L. (1979) The complexity of computing the permanent. Theoret. Comput. Sci., 8, pp. 189–201.

[9] Papadimitriou C.H. (1994) *Computational Complexity* (1st ed.). Addison Wesley. Chapter 9 (NP–complete problems). pp. 181–218.

[10] Jiang J.R. (2008), The theory of NP-completeness. Dept. of Computer Science and Information Engineering. National Central University, Jhongli City, Taiwan.

[11] Plotnikov A.D. (2012) *Heuristic algorithm for finding the maximum independent set.* Cybernetics and Systems Analysis: Volume 48, Issue 5, Page 673-680.

[12] Davis M., Putman H. 1960) A computing procedure for quantification theory // Journal of the ACM. 7 (3). p. 201–215.

[13] Davis M., Logeman G., Loveland D. (1962) A machine program for theorem-proving // Communications of the ACM. 5 (7). p. 394–397.

[14] Kullmann O. (1999) New methods for 3-SAT decision and worst-case analysis // Theoretical Computer Science. 223 (1–2). p. 1–72.